\documentclass{article}
\usepackage{spconf,amsmath,graphicx}
\usepackage[style=numeric, maxnames=1, minnames=1, backend=biber, sorting=none]{biblatex}
\usepackage{enumitem}
\setlist{nosep, leftmargin=14pt}
\usepackage[usenames,dvipsnames]{color}
\usepackage{mwe} 

\newlength{\bibparskip}
\let\oldthebibliography\thebibliography
\renewcommand\thebibliography[1]{%
  \oldthebibliography{#1}%
  \setlength{\parskip}{\bibitemsep}%
  \setlength{\itemsep}{\bibparskip}%
}

\usepackage[skip=1.25pt]{caption}

\usepackage[colorlinks = true,
            linkcolor = blue,
            urlcolor  = blue,
            citecolor = blue,
            anchorcolor = blue]{hyperref}

\title{Optimizing Lymphocyte Detection in Breast Cancer Whole Slide Imaging through Data-Centric Strategies}
\name{
\begin{tabular}{@{}c@{}}
Amine Marzouki \qquad  Zhuxian Guo  \qquad Qinghe Zeng  \qquad  Camille Kurtz \qquad Nicolas Loménie
\end{tabular}}

\address{Laboratoire d'Informatique Paris Descartes (LIPADE), Université Paris Cité, Paris, France \\
}
\begin{document}
%
\maketitle
\begin{abstract}

Efficient and precise quantification of lymphocytes in histopathology slides is imperative for the characterization of the tumor microenvironment and immunotherapy response insights. 
We developed a data-centric optimization pipeline that attain great lymphocyte detection performance using an off-the-shelf YOLOv5 model, without any architectural modifications.
Our contribution that rely on strategic dataset augmentation strategies, includes novel biological upsampling and custom visual cohesion transformations tailored to the unique properties of tissue imagery, and enables to dramatically improve model performances.
Our optimization reveals a pivotal realization: 
given intensive customization, standard computational pathology models can achieve high-capability biomarker development, without increasing the architectural complexity. 
We showcase the interest of this approach in the context of breast cancer where our strategies lead to good lymphocyte detection performances, echoing a broadly impactful paradigm shift. 
Furthermore, our data curation techniques enable crucial histological analysis benchmarks, highlighting improved generalizable potential.

\end{abstract}
\begin{keywords}
Digital pathology, Whole slide imaging, Breast cancer, Lymphocyte detection, YOLOv5, Data-driven approach.
\end{keywords}
\section{Introduction}
\label{sec:intro}

Tumor-infiltrating lymphocytes (TILs) within the tumor microenvironment have demonstrated profound utility as prognostic indicators and in selecting patients likely to benefit from immunotherapies across multiple cancer types \autocite{TILsTumors, TILs}. TILs levels strongly correlate with response to therapies, with high TILs indicating improved prognosis and survival outcomes \autocite{TILsPredictive, TILSurvival}. Beyond prognosis, TILs measures also predict immunotherapy responses, guiding precision medicine treatment decisions \autocite{TILsTumors}. However, efficiently harnessing the power of TILs in the clinical routine remains hampered by time-intensive manual evaluation methods across histopathology slides \autocite{TILsManual}. 

Visual TILs assessment by pathologists is slow, costly, and prone to substantial inter-observer variability that hinders reproducibility \autocite{TILsReproducibility}.
In breast cancer specifically, amidst the heterogeneity across molecular subtypes, Her2 positive and triple negative disease confer more adverse prognoses from their aggressive biological behaviors \autocite{Dai2015BreastCI}. 
Therefore, impactful biomarkers that predict clinical trajectories and guide therapeutic interventions remain imperative clinical needs \autocite{biomarkers}.

Most researches into computational TILs quantification emphasize complex model architectures rather than data curation. 
An example comes from recent work \autocite{Le2020}, who tweaked specialized deep networks like Inception-v4 and 34-layer ResNets in analyzing spatial tumor-TILs distributions. 
Similarly, \autocite{lapp2023pathrtm} developed an advanced object detector called PathRTM to segment individual cell types. While architectural advances have merits, they frequently overshadow the pivotal yet overlooked role of training data itself, which we explore in this article.

As a proof of concept and initial milestone, we leveraged the TIGER \autocite{TigerChallenge} challenge to create a highly optimized lymphocyte detection model for breast cancer H\&E slides. 
By emphasizing extensive dataset augmentation and curation over model modification, our approach achieved great performances. 
This demonstrates the high potential of data-driven optimization for furthering automated computational pathology, even with streamlined models like YOLO \autocite{redmon2016look}.
Additionally, among oncological diseases, colorectal cancer presents a pivotal opportunity to harness TILs assessment for prognosis predictions and guiding immunotherapies \autocite{Immunoscore, ImmunoscoreVal}. 
Optimizing efficient lymphocyte detection pipelines paves the way for automated evaluation directly from histology slides, accelerating development of clinical systems without reliance on immunohistochemistry. 
Overall, our methodology underscores overlooked potential to advance broadly applicable AI-assisted TIL quantification leveraging strategic data enhancements.

\begin{figure*}[!t]
    \centering
    \includegraphics[width=.7\linewidth]{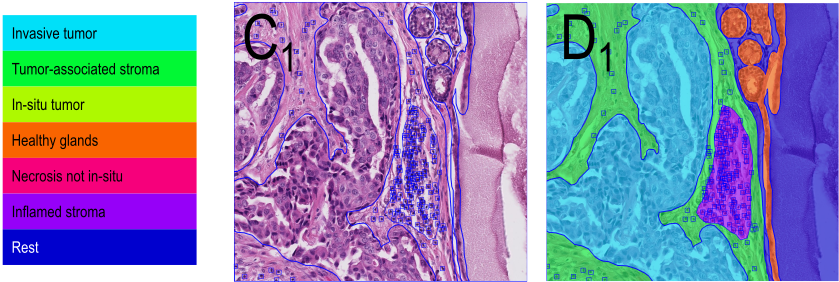}
    \caption{ROI examples from the TIGER challenge dataset \autocite{TigerChallenge}. 
    (C1) tissue borders and lymphocyte bounding boxes / labels. 
    (D1) combination of the tissue compartment masks and bounding boxes.
    }
    \label{fig:annotations}
\end{figure*}

\section{Method}

Object detection models have become ubiquitous in computer vision / medical imaging tasks, providing efficient architectural pipelines for identifying and localizing objects of interest within image contents. 
Popular approaches like single-stage object detection (SSD) frameworks and Faster R-CNN have demonstrated state-of-the-art capabilities across numerous applications.
However, model optimization often dominates attention while tailored data curation remains an underexplored area, especially within computational pathology. 
We hypothesized that by strategically optimizing the training data itself, even streamlined off-the-shelf models could achieve excellent performances without reliance on architectural complexity or domain-specific customization.

After surveying different object detection architectures, we selected the widely adopted YOLOv5 framework for its balance of simplicity and performance. This enabled rapid experimentation with minimal computational overhead during our data-driven exploration. 
Our methodology ultimately centered on intensive dataset optimization rather than extensive model modification or ensemble approaches. 
We sought to test if sufficient data augmentation and curation alone could unlock good lymphocyte detection performances from the streamlined YOLOv5 model.

\subsection{Data}
\label{sec:data}
The data employed in this study was provided in the context of the TIGER challenge \autocite{TigerChallenge} related to the analysis of TILs from H\&E slides for breast cancer.
We used the ROI-level annotations from the provided WSI ROI dataset of tissue-cells, comprising 1879 patches in various sizes with COCO-formatted annotations (Fig.~\ref{fig:annotations}).

\begin{figure}[t]
    \centering
    \includegraphics[width=\linewidth]{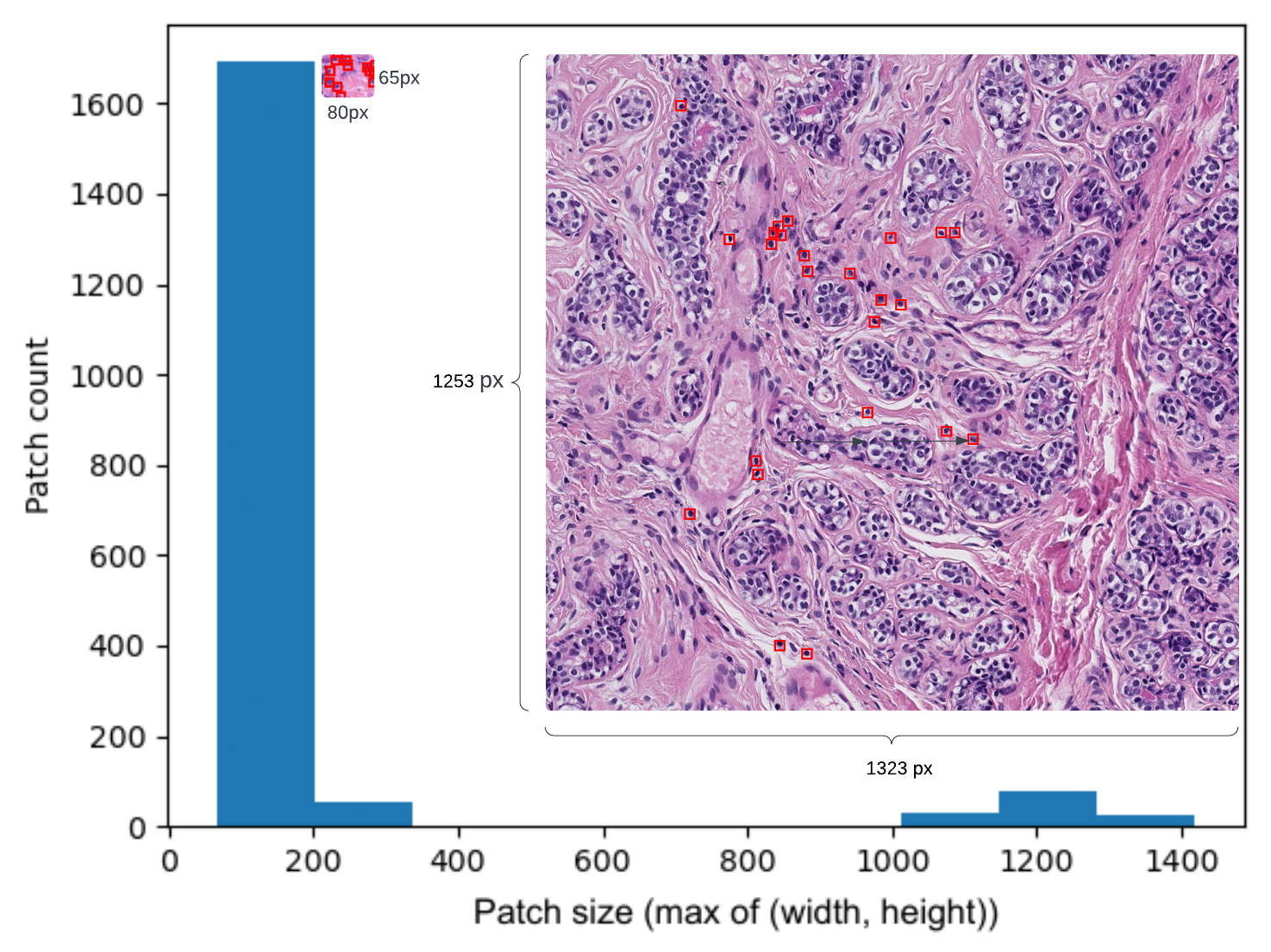}
    \caption{Histogram of the patch sizes. 
    The majority of patches were very small, requiring intelligent upsampling to generate natural 256$\times{}$256 segments. 
    Only a small subset of larger images were tiled into segments retaining full lymphocytes. Tailored upsampling strategies were critical due to the imbalance nature of the dataset.}
    \label{fig:imagesHist}
\end{figure}

A key initial step involved patching the images into 256$\times{}$256 segments, as simple resizing performed poorly. The images exhibited a bimodal size distribution with a large gap between the two subsets (Fig.~\ref{fig:imagesHist}). 
This required tailored patching strategies for each group.

For large patches (subset 1), a tiling approach sliced images into multiple 256$\times{}$256 patches, retaining complete lymphocytes. 
Smaller images (subset 2) required intelligent upsampling to generate natural-looking patches as described hereinafter.

\subsection{Our approach: less on modeling, more on data tailoring}
\label{data_pipeline}

A key innovation of our work relies on considering strategic optimization of the training data itself rather than architectural modifications. 
Our dataset optimization process focused on two main avenues: 1) novel upsampling strategies for small images and 2) visual consistency across patches. 
For upsampling, we generated natural looking expanded patches from smaller ones.
To this end, we employed a multi-faceted stretch strategy including:
\begin{itemize}
\item Mirroring the existing patch to duplicate its content;
\item Extracting random crops from other patches in the dataset and pasting them seamlessly to augment the image size;
\item Isolating and overlaying individual lymphocytes and cells from random patches to enhance biological detail \autocite{ghiasi2021simple};
\item Applying minor augmentation techniques like rotation, blurring and color shifts to smoothly integrate pasted elements except for lymphocytes.
\end{itemize}

By leveraging mirroring, cropping, and lymphocyte transplanting in combination with minor augmentations, the stretch approach organically grew smaller patches to the needed 256$\times{}$256 dimensions. 
The multi-component strategy maintained visual cohesion, increasing plentiful detail while retaining realism. 
This data-driven upsampling proved effective for training the model on properly sized patches derived from tiny source images.

\begin{figure}[t]
    \centering
    \includegraphics[width=0.5\textwidth]{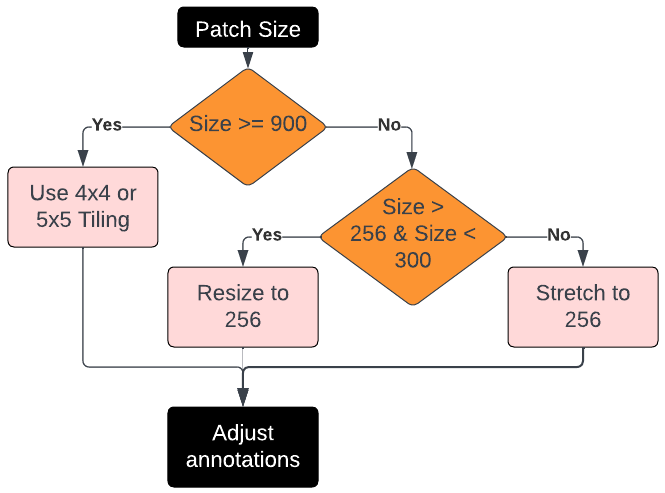}
    \caption{Preprocessing pipeline (the patch size denotes the maximum of patch height and width dimensions). 
    Small patches were upsampled into natural 256$\times{}$256 images via mirroring, cropping, lymphocyte transplanting and minor augmentations.}
    \label{fig:preprocessing}
\end{figure}

For visual consistency, we tailored our patching strategy to retain full lymphocyte bodies and discarded any incomplete segments. Our final curated dataset consisted of two complementary versions: 
one with simple gray padding and one with visually optimized patches. 
Fig.~\ref{fig:preprocessing} summarizes the process.

\section{Results}
\label{sec:results}


YOLOv5 small \autocite{glenn_jocher_2022_6222936} was finetuned for 600 epochs (with a batch size of 32) using default parameters. 
Ultralytics \autocite{Jocher_YOLOv5_by_Ultralytics_2020} native augmentation pipeline provided additional on-the-fly data diversity during training.
Using an RTX 3050 Ti GPU, training completed within 9 hours, achieving 75.77\% precision and 73.03\% recall on the local validation set. 
To refine predictions, we excluded bounding boxes outside 8-20 pixel height/width ranges and adjusted centers of partial boxes. 

The quantitative results of the TIGER challenge are presented in Table~\ref{tab:tiger_ranking}.
The performances of lymphocyte detection are given according to the Free Response Operating Characteristic (FROC) score, computing sensitivity (TPR) versus average false positives (FP) per mm² over all slides.
The YOLOv5 performances boosted from 199th to 6th (2nd place teamwise) after integrating tailored data optimizations, highlighting the impact of the strategic curation embedded in our data-centric optimization pipeline.
The streamlined model design was competitive without architectural elaboration. 


\begin{table}[!t]
 \begin{center}
 \begin{tabular}{lrr}
 \hline
 Method & FROC & TIGER ranking \\
 \hline
 YOLOv5 (w/o) & 0.2139 & 199/329 \\
 YOLOv5 (data-centric optim.) & 0.5731 & 6/329  \\
 \hline
 Biototem  & 0.6003 & 1/329 \\
 TIAger & 0.5718 & 7/329 \\
 \hline
 \end{tabular}
 \caption{Method performances: FROC scores 
 and TIGER challenge ranking comparison with model focused methods.
 }
 \label{tab:tiger_ranking}
 \end{center}
\end{table}

\begin{figure}[!t]
 \centering
  \begin{minipage}[b]{.49\linewidth}
   \centerline{\includegraphics[width=4.0cm]{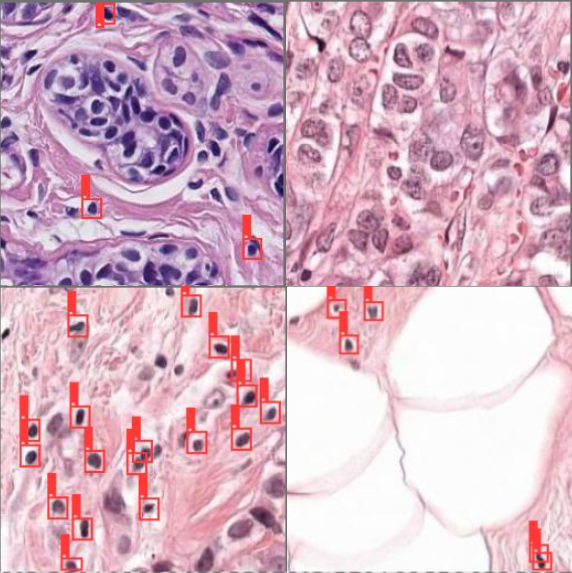}}
   \centerline{(a) ground truth (TIGER)}\medskip
  \end{minipage}
  \hfill
  \begin{minipage}[b]{.49\linewidth}
   \centerline{\includegraphics[width=4.0cm]{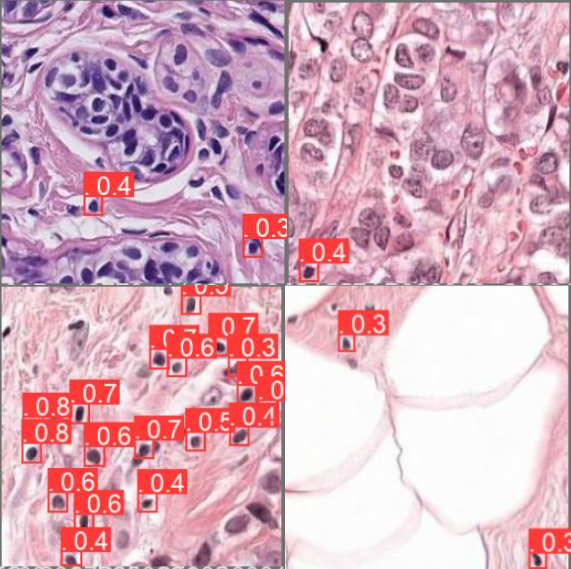}}
   \centerline{(b) model predictions (TIGER)}\medskip
  \end{minipage}
  
  \begin{minipage}[b]{.42\linewidth}
   \centerline{\includegraphics[width=6.5cm]{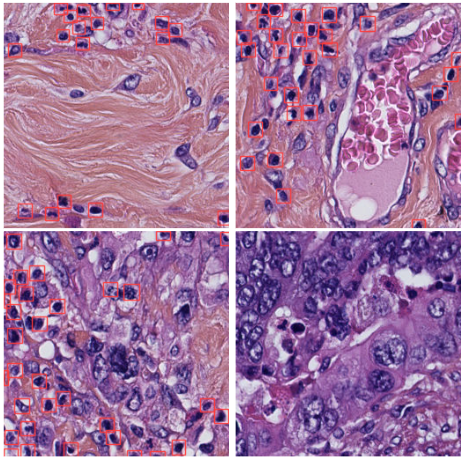}}
   \centerline{(c) inferences on colorectal cancer slides}\medskip
  \end{minipage}
  \hfill
\caption{Validation results on TIGER data and generalization on colorectal cancer slides. 
(a) shows local validation set patches with ground truth lymphocyte annotations. 
(b) displays predictions from our model on the same patches. 
(c) illustrates the TIGER-trained model's original inferences on proprietary colorectal HES slides, demonstrating its potential to be adapted to replace costly IHC methods.}
\label{fig:yolo_result}
\end{figure}

Furthermore, when applied to an in-house colorectal cancer HES dataset (Fig.~\ref{fig:yolo_result}), our lymphocyte detection approach demonstrated strong potential for clinical translation, identifying immune infiltrates with high accuracy without reliance on immunohistochemistry. 



In summary, modest model selection combined with rigorous dataset curation yielded excellent lymphocyte detection capability. Our methodology underscores the immense potential of data-centric techniques to unlock substantial performance gains even with off-the-shelf deep learning models.

\section{Conclusion}
\label{sec:conclusion}
In this work, we presented an efficient data-centric approach for lymphocyte detection in breast cancer histology slides. By strategically optimizing our dataset through novel upsampling techniques and tailored curation, our pipeline attained great performances using an off-the-shelf YOLOv5S model. Competing in the TIGER challenge, our methodology led to a top ranking, demonstrating its potential in automated lymphocyte quantification. 
Overall, intensive dataset customization allowed even a streamlined model to achieve good performance, advancing automated TIL evaluation. 
Our data-centric framework represents a critical step towards reproducible lymphocyte biomarkers to guide prognosis and precision treatment decisions across cancer types. This study highlights the pivotal impact of strategic data curation alongside model optimization to drive the next generation of AI-empowered computational pathology.

%


\section{Compliance with Ethical Standards}
This research study was conducted using data made available in the TIGER challenge that was hosted in grand-challenge.org platform. Ethical approval was not required as confirmed by the license attached with the open access data.

\section{Acknowledgments}
\label{sec:acknowledgments}
This work was supported by Data Intelligence Institute of Paris (diiP), IdEx Université Paris Cité (ANR-18-IDEX-0001) and by the Translational Research Program in Cancerology INCa-DGOS - PRTK-2020. 
It was granted access to the HPC resources of IDRIS made by GENCI.




\printbibliography
\end{document}